\documentclass{aa}  

\usepackage{graphicx}
\usepackage{txfonts}
\newcommand{\RefReply}[1]{{#1}}

\begin{document}

   \title{Suppressing variance in 21-cm signal simulations during reionization}
   % \subtitle{I. Overviewing the $\kappa$-mechanism}

   \author{Sambit K. Giri,\inst{1}\thanks{E-mail: sambit.giri@ics.uzh.ch}
          \and
          Aurel Schneider\inst{1}
          \and 
          Francisco Maion\inst{2}
          \and 
          Raul E. Angulo\inst{2,3}
          }

   \institute{Institute for Computational Science, University of Zurich, Winterthurerstrasse 190, 8057 Zurich, Switzerland.%\\
         \and
    Donostia International Physics Center (DIPC), Paseo Manuel de Lardizabal, 4, 20018 Donostia-San Sebastián, Spain.%\\
        \and 
    IKERBASQUE, Basque Foundation for Science, 48013, Bilbao, Spain.%\\
}

   \date{Received XXX; accepted YYY}

% \abstract{}{}{}{}{} 
% 5 {} token are mandatory
 
  \abstract
  % context heading (optional)
  % {} leave it empty if necessary  
   {
   Current best limits on the 21-cm signal during reionization are provided at large scales ($\gtrsim$100 Mpc). To model these scales, enormous simulation volumes are required which are computationally expensive. 
% Here we study the minimum size of reionization simulations required to analyse current and upcoming observations. We find that the primary source of uncertainty at these large scales is sample variance. 
We find that the primary source of uncertainty at these large scales is sample variance, which decides the minimum size of simulations required to analyse current and upcoming observations.
% In simulations of the large-scale structure, the method of `fixing' the phases of the initial conditions (ICs) to follow a random distribution and `pairing' two simulations with exactly out-of-phase ICs has been shown to significantly reduce sample variance. 
In large-scale structure simulations, the method of `fixing' the initial conditions (ICs) to exactly follow the initial power spectrum and `pairing' two simulations with exactly out-of-phase ICs has been shown to significantly reduce sample variance.
Here we apply this `fixing and pairing' (F\&P) approach to reionization simulations whose clustering signal originates from both density fluctuations and reionization bubbles. Using a semi-numerical code, we show that with the traditional method, simulation boxes of  $L\simeq 500$ (300) Mpc are required to model the large-scale clustering signal at $k$=0.1 Mpc$^{-1}$ with a precision of 5 (10) per cent. Using F\&P, the simulation boxes can be reduced by a factor of 2 to obtain the same precision level. We conclude that the computing costs can be reduced by at least a factor of 4 when using the F\&P approach. 
   }
  % conclusions heading (optional), leave it empty if necessary 
   %{}


   \maketitle
%
%-------------------------------------------------------------------

\section{Introduction}
The 21-cm signal produced by the spin-flip transition of the ground state of neutral hydrogen present during the epoch of reionization (EoR) will be a treasure trove of information. %\citep[e.g.][]{mellema2013reionization}. 
It will not only teach us about the nature of the first luminous sources but also about the thermal and ionization history of the high-redshift ($z\gtrsim 6$) intergalactic medium (IGM).  
See \citet{pritchard201221} for a review.
% Removed \citet{Gnedin:2022eza} as it is about simulation methods, not about 21-cm signal 
Furthermore, it may help reveal the mysteries of the dark matter sector \citep[e.g.][]{Munoz:2018pzp,Schneider2018constraining,Lopez-Honorez:2018ipk,giri2022imprints}, find primordial black holes \citep{Tashiro:2012qe,Mena:2019nhm}, and shed light on the origin of density fluctuations \citep{Furugori:2020jqn,cole2021small}. 

The 21-cm signal can be distinguished from the Rayleigh–Jeans tail of the cosmic microwave background (CMB) radiation using radio telescopes. 
These telescopes will record a quantity known as the differential brightness temperature, which is given by
% \begin{eqnarray}
% \delta T_\mathrm{b} \approx 27 x_\mathrm{HI} (1 + \delta)\left( \frac{1+z}{10} \right)^\frac{1}{2}
% \left( 1 -\frac{T_\mathrm{CMB}(z)}{T_\mathrm{s}} \right)\nonumber\\
% \left(\frac{\Omega_\mathrm{b}}{0.044}\frac{h}{0.7}\right)
% \left(\frac{\Omega_\mathrm{m}}{0.27} \right)^{-\frac{1}{2}} 
% \left(\frac{1-Y_\mathrm{p}}{1-0.248}\right)
% %\left( 1 + \frac{1}{H(z)}\frac{\mathrm{d}v_\|}{\mathrm{d} r_\|}\right)^{-1}
% \mathrm{mK}\ ,
% \label{eq:dTb}
% \end{eqnarray}
\begin{eqnarray}
\delta T_\mathrm{b} \approx 27 x_\mathrm{HI} (1 + \delta)\left(\frac{1+z}{10}\right)^{\frac{1}{2}}
%\left(\frac{0.27}{\Omega_\mathrm{m}}\right)^\frac{1}{2}
\left(\frac{0.15}{\Omega_\mathrm{m}h^2}\right)^\frac{1}{2}
% \left(\frac{\Omega_\mathrm{b}}{0.044}\frac{h}{0.7}\right)
\left(\frac{\Omega_\mathrm{b}h^2}{0.023}\right)
\left( 1 -\frac{T_\mathrm{\gamma}}{T_\mathrm{s}} \right) \mathrm{mK} ,
\label{eq:dTb}
\end{eqnarray}
where $x_\mathrm{HI}$ and $\delta$ are the fraction of neutral hydrogen and the density fluctuation, respectively. Spin temperature $T_\mathrm{s}$ is the excitation temperature of the two spin states of the neutral hydrogen. $T_\mathrm{\gamma}$ is the CMB temperature at redshift $z$. In this work we will assume the spin temperature to be saturated ($T_\mathrm{s}\gg T_\mathrm{\gamma}$), which is a good approximation during the EoR. For simplicity, we furthermore ignore redshift-space distortions (RSD) of the signal. 
% Although RSD have been shown to significantly affect the 21-cm power spectrum in general 
% \citep[e.g.][]{Bharadwaj:2004nr},  
% % \citep[e.g.][]{Bharadwaj:2004nr,ross2021redshift}, 
% they are not relevant for the outcome of this work.

The 21-cm signal from the EoR has not been detected yet. Current radio experiments, such as the 
LOFAR \citep[][]{mertens2020improved}, 
MWA \citep[][]{trott2020deep} %\citep[][]{tingay2013murchison}, 
and HERA \citep[][]{abdurashidova2022first} %\citep[][]{deboer2017hera} 
%Low Frequency Array \citep[LOFAR;][]{mertens2020improved}, the Murchison Widefield Array \citep[MWA;][]{tingay2013murchison}, and the Hydrogen Epoch of Reionization Array \citep[HERA;][]{deboer2017hera} 
provide upper limits of the 21-cm power spectrum  
%\citep{mertens2020improved,trott2020deep,abdurashidova2022first} 
which have helped to rule out some extreme astrophysical models at $z=6-9$ \citep[e.g.][]{ghara2020constraining,ghara2021constraining}. 
% These limits are coming down and in near future, this signal is expected to be detected.
%With lowering of limits from current telescopes and the upcoming powerful Square Kilometre Array \citep[SKA;][]{mellema2013reionization}, this signal is expected to be detected soon.
% A first measurement of the 21-cm clustering signal is expected in the near future \cite{}.
The thermal noise in the 21-cm signal observation increases with wave-mode \citep[e.g.][]{koopmans2015cosmic}. Therefore the best upper limits are currently obtained at wave-modes %scales corresponding to 
$k\sim 0.1\ \mathrm{Mpc}^{-1}$. At even larger scales, the signal cannot be retrieved due to the presence of foreground contamination. Future observations by e.g. HERA and the Square Kilometre Array \citep[SKA;][]{koopmans2015cosmic} are expected to detect the signal at $k\sim 0.1-1$ Mpc$^{-1}$ during their initial observation phases \cite[e.g.][]{greig201521cmmc}. 

\RefReply{The observed field of view (FoV) of radio telescopes are large enough for the observations at $k\gtrsim 0.1$ Mpc$^{-1}$ to be less affected by cosmic variance. For example, the LOFAR upper limits were derived from observations with FoV of $4^{\circ}\times 4^{\circ}$ that corresponds to $k\sim 0.01$ Mpc$^{-1}$ \citep{mertens2020improved}. The error at the scales that we are interested in is dominated by the by various steps in the data processing pipeline, such as calibration and foreground mitigation \citep[e.g.][]{mertens2020improved}. However, the interpretation of these observations will be affected by the variance in simulations that uses a box length ($L$) that is too close to the largest observed scales.}

%Regarding simulations, 
\citet{iliev2014simulating} found that simulations with %box length (L) 
\RefReply{L}$\gtrsim$150 Mpc %of at least $\sim$150 Mpc 
are required to accurately model the distribution and growth of reionization bubbles. However, simulations of this size are known to be strongly affected by sample variance. At the relevant scales ($k\sim 0.1$ Mpc$^{-1}$), the limited number of  wave modes ($N_\mathrm{modes}$) present in a simulation box of $L\sim150$ Mpc results in a sample error that resembles Poisson noise proportional to $ 1/\sqrt{N_\mathrm{modes}}$. %$\propto 1/\sqrt{N_\mathrm{modes}}$
A similar study was done by \citet{kaur2020minimum} including the pre-reionization era.
To reduce sample variance at scales around $k\sim0.1\ \mathrm{Mpc}^{-1}$, huge simulations ($L\gtrsim 500\ \mathrm{Mpc}$) are required \citep[][]{ghara2020constraining}. Alternatively, one can model this signal by averaging over multiple realisations of smaller volume simulations
%it is possible to run multiple realisations of smaller simulation volumes and averaging over the clustering measurements 
\citep[][]{mondal2020tight}. Both approaches are very expensive, especially when considering that the reionization process may be driven by sources residing in dark matter mini-haloes (with masses of $\lesssim 10^8$ M$_\odot$) which need to be properly resolved \cite[e.g.][]{giri2022imprints}. %\citep[e.g.][]{barkana2001beginning}. 

%Recently, \citet{greig2022generating} produced a very large (7.5 Gpc) reionization simulation to study the large-scale modes explored by MWA. Even with a semi-numerical framework, they had to compromise the accuracy to keep the computational expense under control. \textcolor{red}{What do you mean by this? Did they have to go to k-modes smaller than 0.05? Otherwise, what justifies the 7.5 Gpc?} \SG{Greig et al produced these large volumes to study the expected cosmic variance in upcoming observations. They take subvolumes with FoV corresponding to SKA, MWA, etc and estimate the standard deviation. I wanted to make the point that using seminumerical simulation also simulating large-volumes are difficult.} \AS{There is a difference between sample variance (of the sims) and cosmic variance (of the survey). Did Greig study the latter in this work?}

\begin{figure*}
\includegraphics[width=0.98\linewidth]{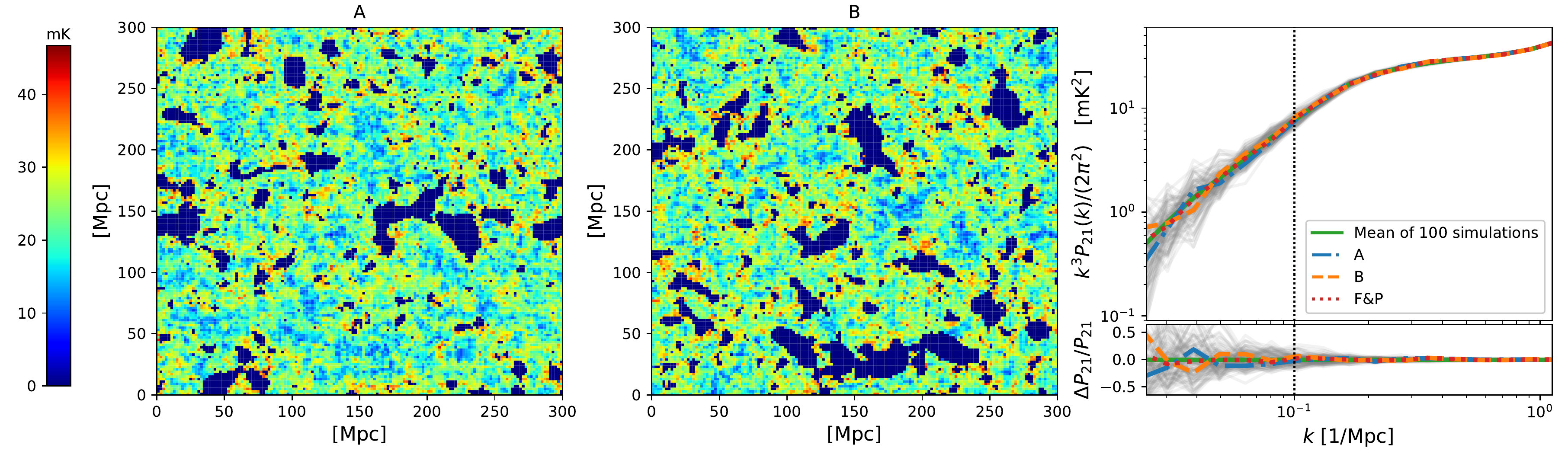}
\caption{{\it First and second panel:} Slices of fixed and paired reionization simulations (A and B) at $z=9$ (with mean neutral fraction of $x_{\rm HI}=0.8$). The colour map shows the differential brightness temperature between 0 and 50 mK. {\it Third panel:} The 21-cm power spectra of the same paired simulations (blue and orange) along with that of %\AS{100 or 300?} 
100 traditional simulations (grey lines). The mean power spectra of the traditional simulations and the 2 F\&P simulations are shown as green solid and red dotted lines, respectively. 
For reference of the expected largest scales probed by SKA, we mark $k=0.1$ Mpc$^{-1}$ with a vertical line.
% We also mark the wavenumbers $k=0.05$ and $0.1$ Mpc$^{-1}$ with vertical lines, which are the largest scales at which the latest upper limits have been put by LOFAR \citep{mertens2020improved} and MWA \citep{trott2020deep} respectively.
%\AS{Why those? you have to mention this here}
}
\label{fig:paired_slices}
\end{figure*}

In this work, we explore a method known as `fixing and pairing' (F\&P) which has been shown to substantially reduce the variance in simulations of matter fluctuations at low redshifts \citep[][]{pontzen2016inverted,angulo2016cosmological}. This method has been extensively used for high-precision predictions of the matter power spectrum \citep[e.g.][]{villaescusa2020quijote,Angulo:2020vky,Euclid:2020rfv}
%, such as the Quijote simultions \citep[][]{villaescusa2020quijote}, %(Quijote sims, Euclid emulator 1,2, Bacco, anything else?) 
as well as biased tracers of the matter distribution \citep[e.g.][]{Villaescusa-Navarro:2018bpd,maion2022statistics}. Here we apply F\&P approach for the first time to simulations of the 21-cm signal during reionization. %which is a highly complex tracer \citep[][]{georgiev2022large}.
%\textcolor{red}{is this true in general? I thought this is not really the case}. \SG{My bad, I meant to say `...which is a complex tracer'.}
In the next section, we describe the simulations used in this study. In Sec. \ref{sec:results}, we present our findings and conclude in Sec. \ref{sec:conclusion}.

% We will implement the F\&P method in reionization simulations. This method constitutes of two parts, which are the following:
% \begin{enumerate}
%     \item Traditionally, the Gaussian initial conditions (ICs) are generated by randomly drawing numbers from a Gaussian zero-mean distribution for both the real and imaginary part of the field in Fourier space. In \textit{fixing}, the ICs are forced to perfectly match the given initial power spectrum. This is achieved by randomly drawing the phases from a flat distribution between 0 and 2$\pi$.
%     \item In \textit{pairing}, two simulations are run with ICs that are exactly out of phase ($\delta^\mathrm{IC}_\mathrm{A}(\mathbf{x})=-\delta^\mathrm{IC}_\mathrm{B}(\mathbf{x})$). 
% \end{enumerate}

% \SG{
% Expand the following:
% \begin{enumerate}
%     \item numbers for the sims required to model observed large scales
%     \item huge volumes or large number of realisations
%     \item improvements in supercomputers and observing precision
%     \item define the 21-cm signal and $\delta T_\mathrm{b}$ here
% \end{enumerate}
% }

\section{Simulations}

For this work, we use the publicly available reionization simulation code \textsc{21cmFAST} \citep[][]{mesinger201121cmfast}. %\citep[][]{murray202021cmfast,mesinger201121cmfast}. 
We modified the initial condition (IC) generator of the code, adding the option of fixing the IC. The original (or Gaussian) method of {\texttt 21cmFAST} is summarised in Sec.~\ref{sec:sim_trad} and our modifications are discussed in Sec. \ref{sec:sim_FnP}.

\subsection{Gaussian method}
\label{sec:sim_trad}

\textsc{21cmFAST} initializes a Gaussian random field (GRF) at the initial redshift ($z_\mathrm{init}=50$) using a linear power spectrum $P(k,z_\mathrm{init})$ obtained from the \citet{eisenstein1999power} fitting function. The ICs of the density perturbations are given by
\begin{eqnarray}
\delta(\textbf{k},z_\mathrm{init}) = \sqrt{0.5P(k,z_\mathrm{init})}~ [a_\textbf{k}+ib_\textbf{k}] \ ,
\label{eq:IC_trad}
\end{eqnarray}
where $a_\textbf{k}$ and $b_\textbf{k}$ are drawn from a Gaussian distribution $\mathcal{N}(0,1)$ \citep{mesinger2007efficient}.
The subsequent formation and evolution of structures are simulated using second-order Lagrangian perturbation theory \citep[2LPT; e.g.][]{bernardeau2002large}. The growth of ionized bubbles during reionization is modelled based on the formalism in \citet{furlanetto2004growth}.

% To study the variance of the large-scale statistics, we use three different box lengths that 200, 300 and 400 Mpc. All these simulations have a spatial resolution of 2 Mpc. 
%The topology of reionization bubbles is caused by the nature, the abundance, and distribution of luminous sources \citep[e.g.][]{giri2021measuring}. \textcolor{red}{I don't understand what you mean with this.}
% In this work, we are not interested in exploring various reionization models. %Therefore we set all the astrophysical parameters in \textsc{21cmFAST} to the fiducial model in \citet{Greig201521CMMC:Signal} that is used to create the mock observation. The minimum mass of the dark matter haloes containing luminous sources is $10^8$
% Therefore we fix the astrophysical parameters to fiducial values given in table 1 of \citet{park2019inferring}.
% %Table~\ref{tab:param_values}. 
% See \citet{park2019inferring} for detailed description of these astrophysical parameters. Note that in this study we will focus on reionization epochs.
In this work, we assume the default values provided in \citet{greig201521cmmc} for the source parametrization (model 1). The virial temperature of the smallest haloes containing stars is set to $T_\mathrm{vir}\simeq 10^{4.7}$ K. %$T_\mathrm{vir}=5\times 10^4$ K. 
For simplicity, the ionising efficiency of sources is set to $\zeta=20$. 
Note that in reality $\zeta$ is expected to scale with source properties, e.g. with the mass of the hosting halo \citep[see e.g.][]{park2019inferring,schneider2021halo}. 
Finally, the maximum distance that photons can travel is set to $R_\mathrm{max}=15$ Mpc, which models the unresolved absorbers. See \citet{georgiev2022large} for more discussion. %For a more realistic modelling of inhomogeneous recombination, see \citet{Sobacchi:2014rua}.

Next to the fiducial model 1, we investigate two more models where reionization is caused by rare and very efficient sources (model 2) and by a larger number of inefficient sources (model 3). The parameters of model 2 and 3 are given by  
$\{T_\mathrm{vir},\zeta\} = \{ 10^{5.5}\mathrm{K}, 200\}$ and $\{T_\mathrm{vir},\zeta\} = \{ 10^{4.3}\mathrm{K}, 15\}$, respectively.
% $\{T_\mathrm{vir},\zeta\} = \{ 3\times 10^5\mathrm{K}, 200\}$ and $\{T_\mathrm{vir},\zeta\} = \{ 2\times 10^4\mathrm{K}, 15\}$, respectively.
% $\{T_\mathrm{vir},\zeta,R_\mathrm{max}\} = \{ 3\times 10^5\mathrm{K}, 200, 15\mathrm{Mpc}\}$ and $\{T_\mathrm{vir},\zeta,R_\mathrm{max}\} = \{ 2\times 10^4\mathrm{K}, 15, 15\mathrm{Mpc}\}$, respectively. 
Note that the parameters of all three models are chosen such that the reionization history remains unchanged.
To study the influence of sample variance, we run simulations with three different box lengths of $L=100$, 150, 200, 300 and 400 Mpc fixing the spatial resolution to 2 Mpc.

\subsection{Fixing and pairing method}
\label{sec:sim_FnP}

The F\&P method for simulations is a two step process that was introduced in \citet{angulo2016cosmological}. The first step consists of \textit{fixing} the ICs, 
%This means that the initial density fluctuations are not obtained via Eq.~\ref{eq:IC_trad} but rather by using
and is achieved by replacing Eq.~\ref{eq:IC_trad} with the following,
\begin{eqnarray}
\delta(\textbf{k},z_\mathrm{init})=\sqrt{P(k,z_\mathrm{init})} \exp(i\theta_\textbf{k}).
\end{eqnarray}
% As a consequence, instead of drawing two numbers ($a_{\mathbf{k}}$, $b_{\mathbf{k}}$) from a Gaussian distribution, only one number ($\theta_{\textbf{k}}$) has to be randomly drawn 
Here we randomly draw $\theta_{\textbf{k}}$ from a flat distribution between 0 and 2$\pi$. 
This way the IC is \textit{fixed} to exactly give the linear power spectrum.
% The real and imaginary part of the phases are thereby \textt{fixed}. 
The second step of the F\&P method consists of running two \textit{paired} simulations (A and B) that are identical except that the phases of B are shifted of $\pi$ with respect to those of A. This means that their density fields are inversed, i.e. $\delta_\mathrm{A}(\textbf{k},z_\mathrm{init})=-\delta_\mathrm{B}(\textbf{k},z_\mathrm{init})$. %The gain of the F\&P method is that both steps, 
Fixing the ICs and taking the average over summary statistics, such as power and bi-spectra, of two paired simulations significantly reduces the sample variance. %when measuring. %\citep{angulo2016cosmological}. 
%e.g. the power or bi-spectra of the matter distribution \citep{angulo2016cosmological} or other biased tracers \citep{Villaescusa-Navarro:2018bpd}. The goal of the present work is to investigate whether the F\&P method remains successful for simulations of the reionization process.

In the first two panels of Fig.~\ref{fig:paired_slices}, we show slices of the 21-cm signal from two \textit{paired} simulations produced with the F\&P method. The slices are shown at $z=9$ which corresponds to an early phase of reionization where the majority of the simulation volumes are still neutral (mean neutral fraction of $x_\mathrm{HI}=0.8$). Here the bubbles are visible as dark blue areas. Their positions are strongly correlated with the matter perturbations of the corresponding simulation. As the initial density fields between the two simulations are perfectly anti-correlated, the bubbles are anti-correlated as well. The two slices confirm that ionised regions in one simulation are still neutral in the other. Note,
however, that this anti-correlation, while being very strong during the early phases of reionizations, reduces with time as the bubbles grow larger and merge. %\citep[][]{furlanetto2016reionization}. %to form the \textit{percolation cluster} \citep[][]{furlanetto2016reionization}. %We will discuss more about this feature in the following sections.

The third panel of Fig.~\ref{fig:paired_slices} shows the corresponding power spectra from the paired simulations (A and B) in blue and orange. The resulting F\&P power spectrum, corresponding to the mean of the two, is shown in red. For comparison, we also provide the mean power spectrum of 100 independent realisations with the Gaussian method (green line). %same volume (green line). 
The individual power spectra from each of these simulations are shown in grey. The fact that the F\&P result (red) is nearly indistinguishable from this mean power spectrum (green) %of 100 independent simulations (green) 
is very promising. It qualitatively confirms that the F\&P method can reduce sample variance for simulations of the EoR. In the following section, we will investigate this result in a more quantitative manner. %This is a non-trivial result because reionization bubbles are not simple biased tracers of the underlying density field. 

\section{Results}
\label{sec:results}
In this section, we will first show that sample variance is indeed the dominating modelling error for reionization simulations. We then investigate how we can reduce the sample variance using the F\&P method and  provide estimates for the smallest volumes that can be run without being dominated by sample variance.

\subsection{Power spectrum}
\label{sec:result_ps} % used for referring to this section from elsewhere

\begin{figure}
\includegraphics[width=0.88\linewidth]{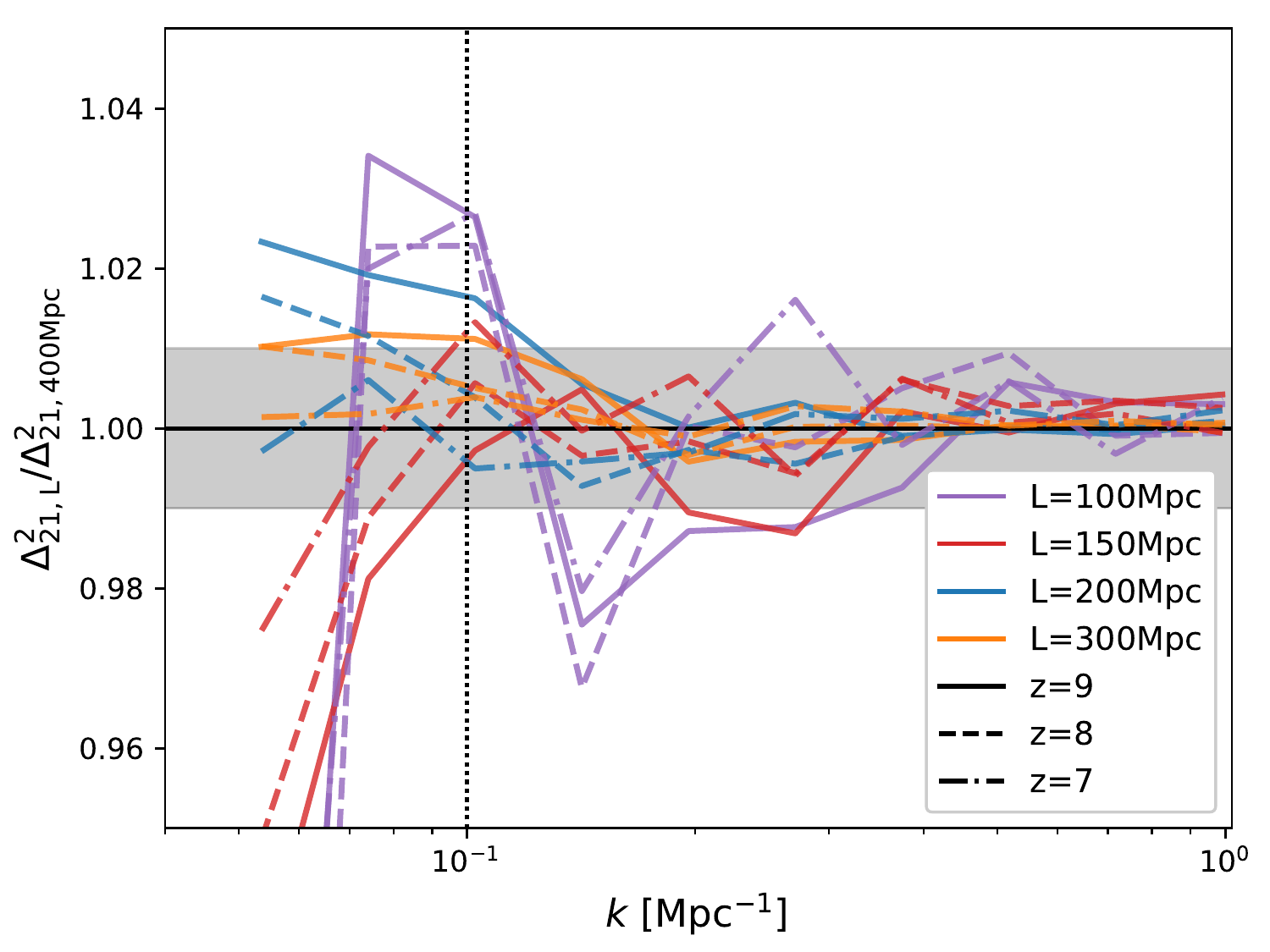}
\caption{The ratio of the 
power spectra produced with small simulation volumes ($L=100$, 150, 200 and 300 Mpc) to that with the largest simulated volume ($L=400$ Mpc). 
% We also mark the wavenumbers $k=0.05$ and $0.1$ Mpc$^{-1}$ with vertical black dotted lines. 
We mark $k=0.1$ Mpc$^{-1}$ with a vertical line and 1 per cent level with grey shaded region. 
The differences at $k\gtrsim 0.1$ Mpc$^{-1}$ are within per cent level for simulations with $L\gtrsim 150$ Mpc.
}
\label{fig:result_ps_vs_L}
\end{figure}

\begin{figure*}
\centering
\includegraphics[width=0.93\linewidth]{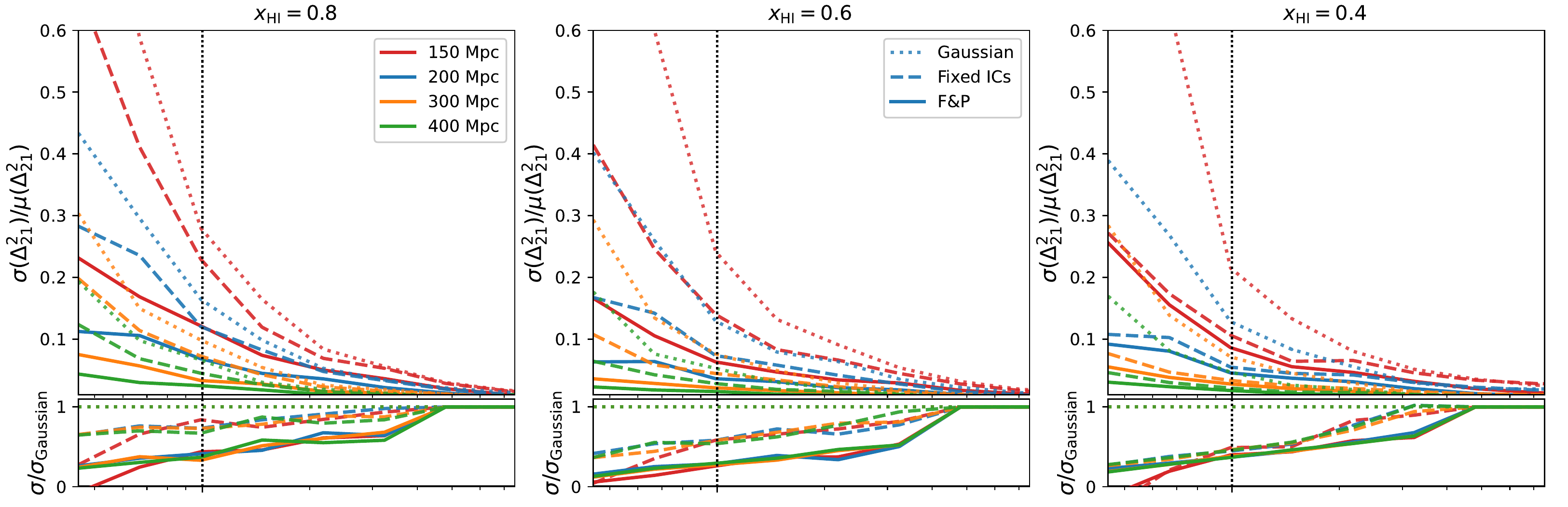}
\includegraphics[width=0.93\linewidth]{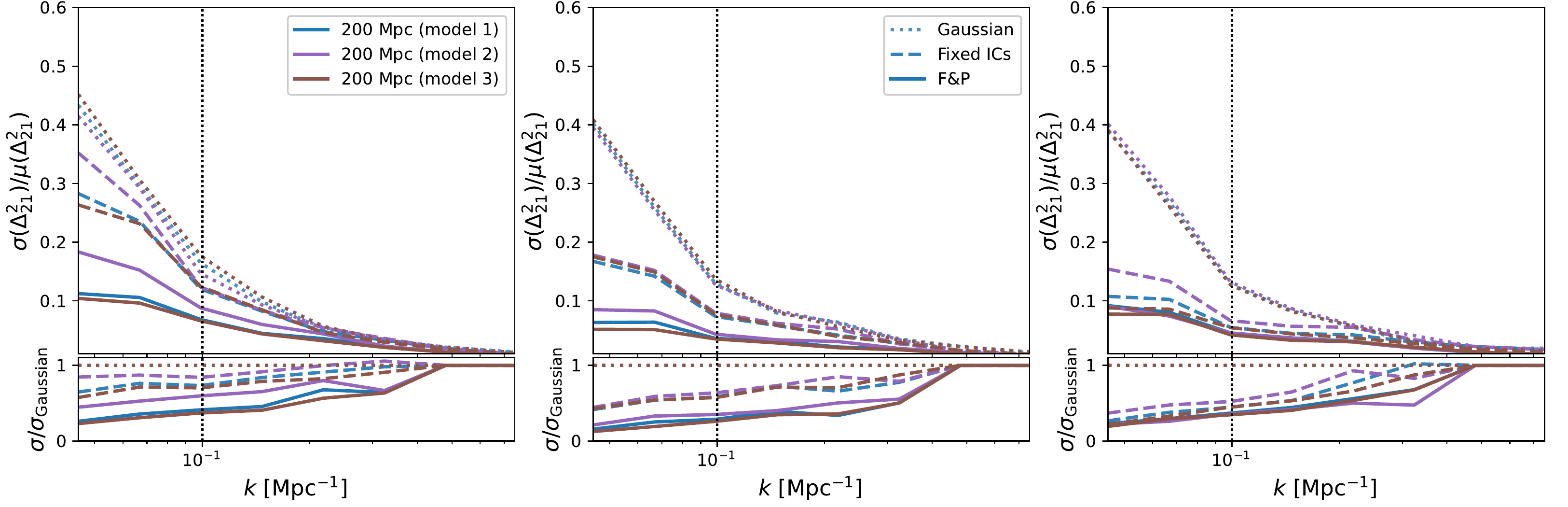}
\caption{The ratio of standard deviation on the power spectra to the mean power spectra estimated from 100 simulations at early (left-panels), middle (middle-panels) and late (right-panels) stages of reionization.
We also mark the wave mode $k=0.1$ Mpc$^{-1}$ with vertical lines.
The top panels show the results from four different simulation volumes (150 Mpc: red, 200 Mpc: blue, 300 Mpc: orange, 400 Mpc: green) for the Gaussian (dotted), fixed ICs (dashed) and F\&P (solid) method. We see that fixed ICs and F\&P methods reduce the error on the power spectra by about 1.5 and 2 times respectively compared to the Gaussian method.
The bottom panels show the results for three different reionization models (1: blue, 2: violet, 3: brown). 
Both fixed ICs and F\&P methods work in a similar manner for all the models. 
We have also included bottom sub-panels showing the ratio of standard deviation to that of the Gaussian method.
}
\label{fig:result_ps}
\end{figure*}

The power spectrum is expected to be the first detectable statistics of the 21-cm signal from radio interferometric experiments. We therefore require accurate predictions of the power spectrum with an associated theoretical error that ideally stays below the observational errors for all wave modes and redshifts. In terms of simulations, this means that we have to quantify errors related to e.g. the resolution, box-size, and sample variance, selecting a simulation setup where such errors are sub-dominant. In this work, we aim to quantify the minimum box-size required for reionization simulations. This means we do not investigate resolution effects but we focus on errors caused by missing large-scale modes and sample variance.

In Fig.~\ref{fig:result_ps_vs_L}, we show the ratio of the power spectra of smaller volumes compared to our largest simulation ($L=400$ Mpc). Since these simulations are set up with the same realisation of the density field, any deviations of smaller simulations are caused by missing modes that are larger than the box size. %In the relevant regime of 
At $k\gtrsim 0.1$ Mpc$^{-1}$, we find that the power spectra from small volume simulations ($L=150$ (red), 200 (blue), and 300 Mpc (orange)) deviate at a per cent level from the power spectrum measured on the $L=400$ Mpc box. This is true for all redshifts investigated. 
% Only at very large scales ($k=0.05$ Mpc$^{-1}$) the smallest simulation ($L=150$ Mpc) starts to deviate substantially (by $\sim$5 per cent) while the other simulations remain within the 1 per cent level. This deviation is likely due to a Poisson noise contribution given that the largest scales are estimated with only $N_\mathrm{modes}=12$ modes. % Removed this as we are not focusing on k=0.05 anymore.
We also plot the case with $L=100$ Mpc (violet). This case shows large deviations at most wave modes as this volume is affected by the missing large-scale modes, which is consistent with \citet{iliev2014simulating}.
Since the upcoming observations from the SKA are expected to provide measurements 
at $k\gtrsim 0.1$ Mpc$^{-1}$ \cite[e.g.][]{greig201521cmmc}, 
% within the range of $k\sim 0.1-1$ Mpc$^{-1}$ \cite[][]{greig201521cmmc}, 
we conclude from Fig.~\ref{fig:result_ps_vs_L} that a box size of $L=150$ Mpc is sufficient to model the largest required modes %to capture the largest modes affecting the signal 
at the per cent level. 
See \citet{greig2022generating} for a similar study.

We now turn our attention to the error caused by sample variance. The top panels of Fig.~\ref{fig:result_ps} show the standard deviations ($\sigma$) with respect to the mean power spectrum ($\mu$) from 100 simulations at three different redshifts (corresponding to $x_{\rm HI}=0.8$, 0.6, and 0.4 from left to right). %This means that out of the 100 simulations, 32 deviate by more than the value provided by the corresponding line at a given $k$-value. 
The sample variances obtained with the Gaussian method (Sec.~\ref{sec:sim_trad}) %(i.e. using ICs with different random seeds) 
are shown as dotted lines, where different colours correspond to different simulation volumes ($L=150$, 200, 300, and 400 Mpc in red, blue, orange, and green, respectively). The results from fixed simulations and F\&P simulations are shown as dashed and solid lines with the same colour schemes. 
While already fixing the ICs helps to reduce the sample variance, a more significant improvement is obtained with the F\&P method. Independently of the simulation volume and the $k$-mode, F\&P leads to a suppression of the cosmic variance by about a factor of $\gtrsim$2 compared to the Gaussian method. 

In the bottom panels of Fig.~\ref{fig:result_ps}, we focus on the $L=200$ Mpc box providing results from different astrophysical models 1 (blue), 2 (purple), and 3 (brown). All three models provide a similar improvement, confirming that our results are only weakly dependent on the choice of astrophysical model. %Note, however, that the improvement obtained by the F\&P method is slightly worse for model B. This situation occurs as the topology of the 21-cm signal comprises of very large and rare ionised bubbles compared to the other two models. These large bubbles can overlap more and therefore loose more of the anti-correlated regions in the 21-cm signal distribution. However, we still see substantial reduction in the standard deviation for model B compared to the traditional simulation method, which indicates that the suppression of error with the F\&P method in quite generic. In the future, we plan to carry out a more complete parameter study with this method.

Comparing different simulation volumes shown in Fig.~\ref{fig:result_ps}, we conclude that the F\&P method can obtain the same sample variance for simulations that have a 2 times smaller box-size than with the traditional approach. For example at $k=0.1$ Mpc$^{-1}$, the F\&P approach gives us similar sample variance for the $L=200$ Mpc simulations compared to the $L=400$ Mpc simulations with the Gaussian method at all redshifts.
%leads to a 5-10 percent sample variance for a simulation volume with $L=200$ Mpc, the exact value depending on the redshift and the astrophysical model. Using the traditional method a similar sample variance is obtained with a box length that is twice as large (i.e. $L=400$ Mpc). 
Reducing the box-size by a factor of 2 improves the speed and reduces the memory requirement by at least a factor of 8. Since 2 simulations are required for the F\&P approach a gain of at least a factor of four is expected.

\subsection{Bispectrum}
\label{sec:result_bispec}

\begin{figure}
\includegraphics[width=0.9\linewidth]{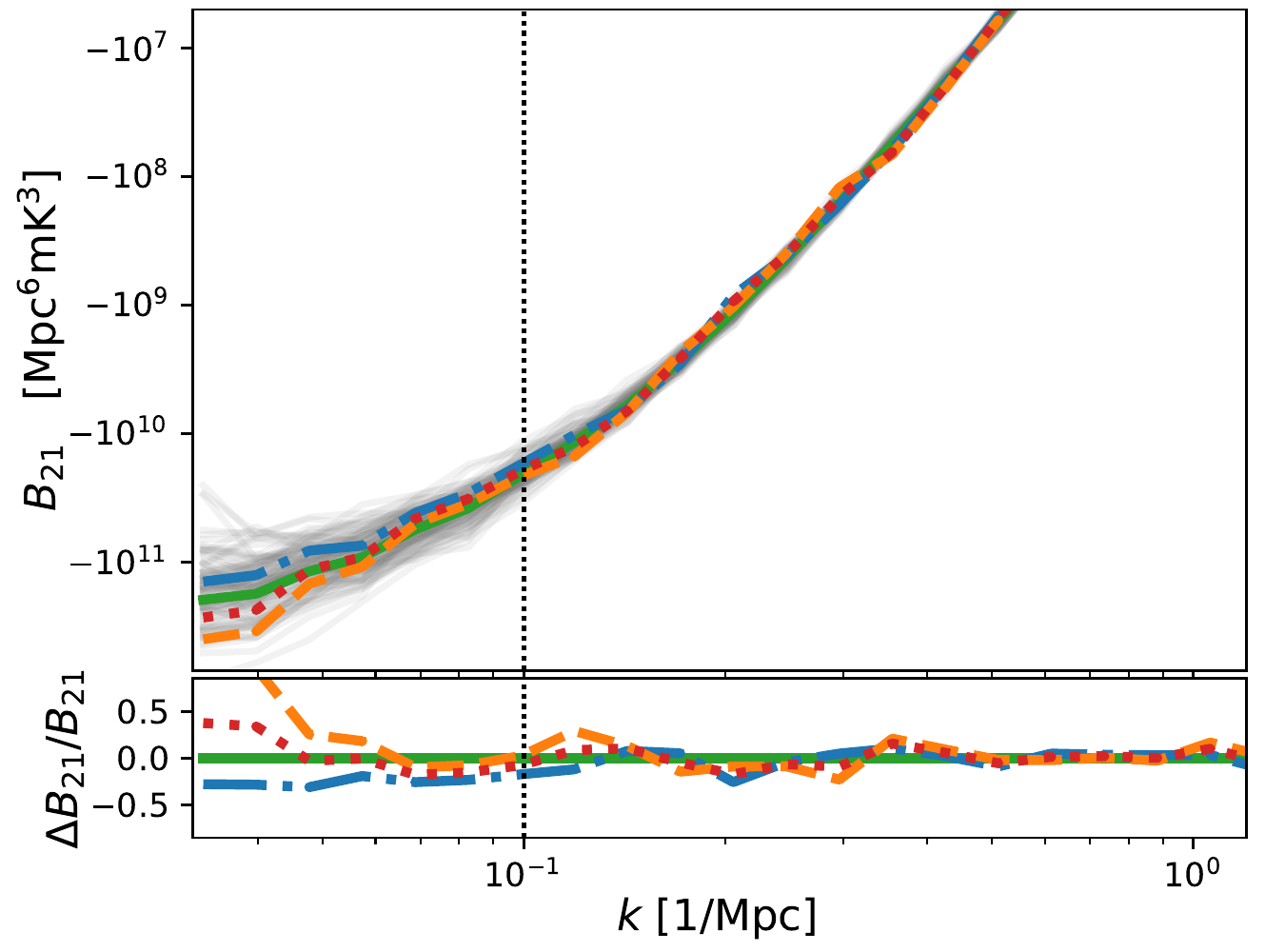}\\
\includegraphics[width=0.9\linewidth]{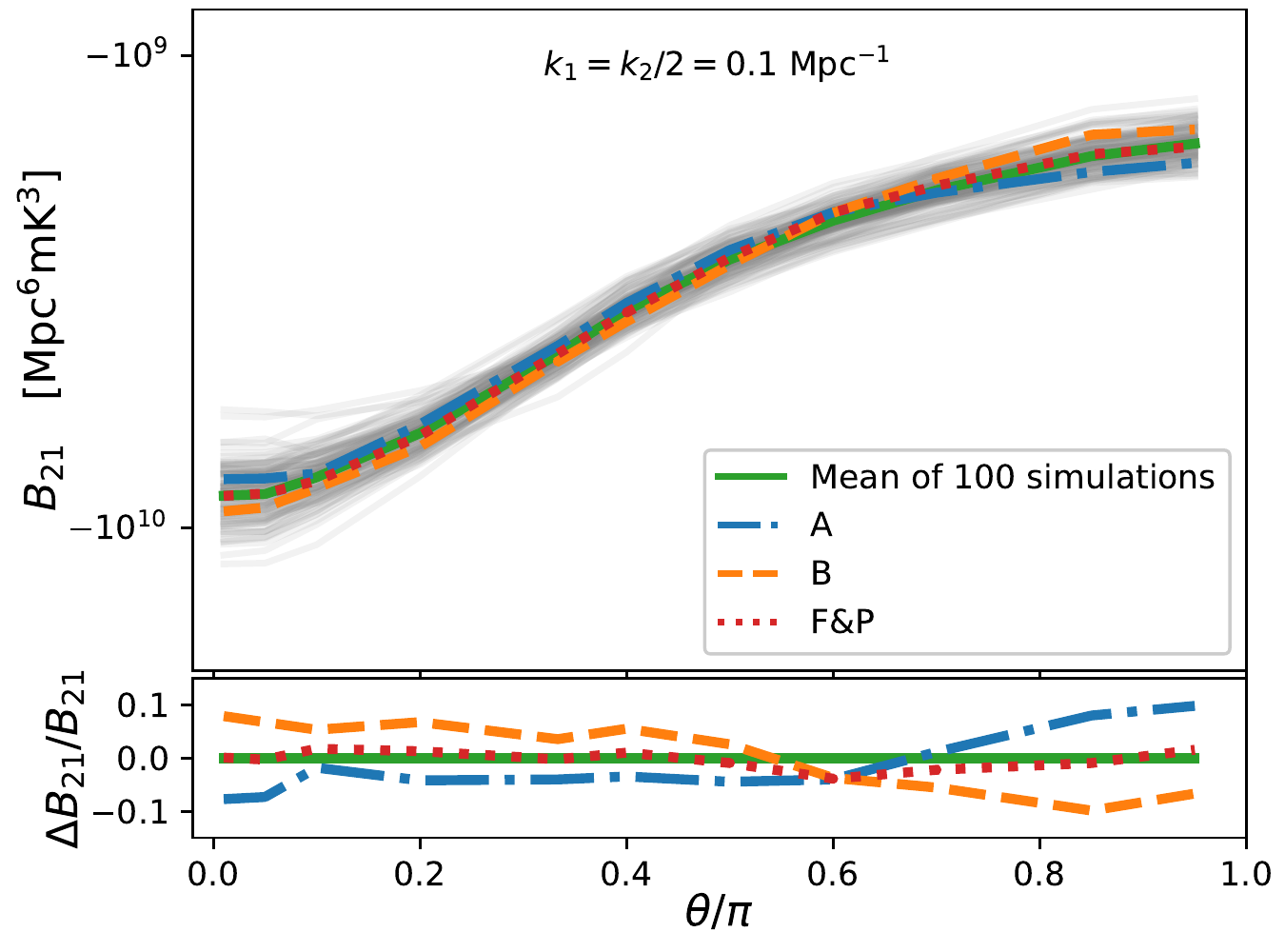}
\caption{The 21-cm bispectra from 100 simulations at $z=9$ produced in $L=300$ Mpc simulation volumes is plotted with grey lines and the mean is given with green lines in both panels.
% The mean bisepctra of these 100 simulations are given with  green solid lines in both panels. 
The top panel shows the equilateral bispectra as a function of wave modes.
We also mark the $k=0.1$ Mpc$^{-1}$ with a vertical lines and observe that the variance increases below this scale.
The bottom panel shows a scalene bispectra as a function of the angle between the two wave vectors $\mathbf{k}_1$ and $\mathbf{k}_2$. 
The 21-cm bispectra corresponding to the simulation A and B shown in Fig. \ref{fig:paired_slices} are given with blue and orange lines respectively. With these 2 simulations, we can estimate the bispectra (red lines) that is very close to the mean bispectra in both panels.
}
\label{fig:result_bispec}
\end{figure}

The 21-cm signal is expected to be highly non-Gaussian. As a consequence, higher-order statistics will have to be used to obtain all the available information contained in the 21-cm density field. Measuring the bispectrum $\mathcal{B}$ consists of an obvious step in that direction \citep[e.g.][]{majumdar2018quantifying}, which is given by%. The 21-cm bispectrum $B$ is given by
\begin{eqnarray}
\left< \delta T_\mathrm{b}(\mathbf{k}_1) \delta T_\mathrm{b}(\mathbf{k}_2) \delta T_\mathrm{b}(\mathbf{k}_3) \right> = \delta_D(\mathbf{k}_1+\mathbf{k}_2+\mathbf{k}_3) \mathcal{B}(\mathbf{k}_1, \mathbf{k}_2, \mathbf{k}_3) \ ,
\label{eq:bispec_def}
\end{eqnarray}
where $\delta_D$ is the Dirac delta function. $B(\mathbf{k}_1, \mathbf{k}_2, \mathbf{k}_3)$ depends on the configuration of triangles formed by the three wave-vectors ($\mathbf{k_1}$, $\mathbf{k_2}$, and $\mathbf{k_3}$). Here we study two configurations which are the equilateral ($k_1=k_2=k_3$) and a scalene ($k_1=k_2/2=0.1~\mathrm{Mpc}^{-1}$) triangle. We use the publicly available package \textsc{BiFFT} \citep[][]{watkinson2021bifft}
%\footnote{\url{https://bitbucket.org/caw11/bifft/src/master/}} \citep[][]{watkinson2017fast} 
to measure the bispectra of our simulations.

% In the bottom-middle panel of Fig.~\ref{fig:paired_slices}, show the equilateral bispectrum $B(k)$. We see similar behaviour as the the power spectrum. Based on \citet{trott2019gridded}, we focus on $k=0.1$ Mpc$^{-1}$ for the $B(k)$. The F\&P method again helps in modelling the true $B(k)$ at large-scales.

% The top and bottom panels of Fig. \ref{fig:result_bispec} show the equilateral and scalene bispectra respectively at $z=9$. 
The top panel of Fig. \ref{fig:result_bispec} shows the equilateral $\mathcal{B}$ at $z=9$ ($x_{\rm HI}=0.8$) for the simulations shown in Fig,~\ref{fig:paired_slices}. The 100 simulations with the Gaussian method are plotted in grey (with their mean in green) and the simulations A and B are highlighted in blue and orange. Just as for the power spectrum, we observe significant sample variance in the equilateral $\mathcal{B}$ at small wave modes ($k\lesssim 0.2$ Mpc$^{-1}$). For reference, we again mark the $k=0.1$ Mpc$^{-1}$. %, which is the scale where the $\mathcal{B}$ is expected to be first detected \citep{trott2019gridded}. % This paper is a bit confusing. It doesn't really say that this is the best scale to observe B.
Finally, the equilateral $\mathcal{B}$ of the F\&P simulations is shown in red. It lies very close to the mean value of the 100 independent simulations, confirming the results obtained with the power spectrum. As a consequence, we expect the F\&P method to yield a similar improvement regarding the sample variance of the equilateral $\mathcal{B}$. 

Very similar conclusions can be drawn when investigating the scalene $\mathcal{B}$ shown in the bottom panel of Fig. \ref{fig:result_bispec}. Here the $\mathcal{B}$ is given as a function of the opening angle between the two wave vectors $\textbf{k}_1$ and $\textbf{k}_2$ ($\theta = \mathrm{cos}^{-1} (\textbf{k}_1.\textbf{k}_2/(k_1k_2)$).  %($\theta = \mathrm{cos}^{-1} (\frac{\textbf{k}_1.\textbf{k}_2}{k_1k_2})$). 
As $k_1=k_2/2$ = 0.1 Mpc$^{-1}$, this $\mathcal{B}$ is probing the non-Gaussianity at large scales for all values of $\theta$. Therefore, we observe large variance for all values of the angle $\theta$. We again find that the F\&P simulations help %with the F\&P method helps 
in getting close to the mean $\mathcal{B}$ estimated from 100 traditional simulations. 
In this section we have argued that the sample variance of the $\mathcal{B}$ can be suppressed by the F\&P method in a similar way as for the power spectrum. We limited ourselves to a qualitative analysis of two particular $\mathcal{B}$ configurations for a single redshift. Note, however, that the non-Gaussian information contained in the 21-cm bispectra is very rich, showing a complicated evolution during the EoR \citep[][]{majumdar2018quantifying}. % A detailed study is beyond the scope of this work. 
A more thorough investigation of the effect of sample variance on higher-order statistics is left for future work.

\section{Conclusions}
\label{sec:conclusion}
Numerical simulations of the EoR are very expensive as they need to simultaneously resolve small sources and cover large cosmological volumes. In this work, we use the semi-numerical code {\tt 21cmFAST} to investigate the minimum box-size ($L$) a simulation needs to produce unbiased results. We thereby primarily focus on the 21-cm power spectrum at scales corresponding to $k=0.1-1\ \mathrm{Mpc}^{-1}$, where future observations from e.g. HERA and SKA are expected to detect %provide first detection of 
the 21-cm signal. 

First, we performed a comparison of power spectra from simulations with the same initial density field but different box lengths ($L=100$, 150, 200, 300, and 400 Mpc). The analysis revealed that power spectra from $L\gtrsim 150$ Mpc agree within a per cent in the regime of $k=0.1-1\ \mathrm{Mpc}^{-1}$. We conclude that a box-size of $L=150$ Mpc is sufficient to be unbiased by missing large-scale modes %exceeding the simulation box.

We then studied sample variance (sometimes referred to as cosmic variance) which is known to affect the large-scale power spectrum at a more significant level. Using the Gaussian method, we show that simulation volumes with a box-length of at least $L=400$ Mpc are required to reduce the uncertainty to less than 10 percent at $k= 0.1\ \mathrm{Mpc}^{-1}$.
The sample variance can be reduced by about a factor of 1.5 by \textit{fixing} the ICs. With this method, we can achieve  $\lesssim 10$ per cent error at $k \approx 0.1\ \mathrm{Mpc}^{-1}$ by using a smaller simulation box of $L=300$ Mpc instead. Note that the reionization simulation of the THESAN project \citep[][]{kannan2022introducing} were fixed, but they did not study impact of `pairing'.

As a further step, we apply the F\&P method to reionization simulations. The method consists of taking the average power spectrum from 2 fixed simulations with inverted ICs. While F\&P simulations have been successfully used to suppress sample variance for low-redshift cosmological simulations, they have never been used in the context of reionization. We show that the F\&P method can further reduce the effect of sample variance to below 10 per cent for a box-size of $L=200$ Mpc. 
We also tested the robustness of our results by changing the astrophysical parameters assuming a model with fewer, more efficient sources and a model with a larger number of inefficient sources. %that all lead to the same global ionisation history but different power spectra. 
% We check that these models only lead to small changes in the sample variance, 
We found similar improvement these models,
which means that our general conclusions remain independent of the assumed astrophysical model.

Finally, we investigated the effect of sample variance on the 21-cm bispectrum. We found that the bispectrum is similarly affected by sample variance as the power spectrum. The F\&P method is expected to improve the theoretical predictions for higher-order statistics as well, in agreement with findings from cosmological simulations at low redshifts \citep[][]{angulo2016cosmological}.

Note that we have not included RSD in our simulation. We do not expect significant impact on our findings as to the first order, the RSD just boosts the signal at all wave modes \citep[e.g.][]{ross2021redshift} % \citep[e.g.][]{Bharadwaj:2004nr} 
which will cancel out in the relative error studied here. In the future, we will explore this effect in detail.

\RefReply{}

In general, we conclude that the F\&P method allows to significantly reduce sample variance caused by the finite simulation volume. For reionization, this means that two F\&P simulations with $L=200$ Mpc are sufficient to predict the 21-cm signal in the regime of $k=0.1-1$ Mpc$^{-1}$ to better than 10 per cent. This give us of a speed up of a factor of $\gtrsim$4 compared to the previous simulation method.
%This consists of a speed up of at least a factor of four compared to the traditional simulation method.

%In this work, we used a semi-numerical code to study the variance in the modelled signal as simulating many realisations with a numerical reionization simulation will be computationally expensive. However, our findings would be similar for a numerical simulations as previous studies have shown that they agree at large scales \citep[e.g.][]{majumdar2014use}. Though the suppression of variance is less in \textit{fixed} IC simulations compared to the \textit{paired} ones, it is a useful method to remember while using a very expensive simulation method that cannot be run twice.

%One caveat of this analysis is that we have not considered the super-sample bias due to which the patch of the sky used to estimate the power spectrum or the bispectrum would not have a flat cosmology \citep[e.g.][]{wagner2015separate}. Hence the reionization epoch in the small patch would be slightly different from the global case \citep{giri2019position}. \citet{giri2019position} showed that the power spectra will be affected at all scales and the strength of this effect is sensitive to the source model and reionization history. We will study this effect in more detail in the future. 

\begin{acknowledgements}
We thank Bradley Greig for useful comments.
This research was supported by the Munich Institute for Astro-, Particle and BioPhysics (MIAPbP) which is funded by the Deutsche Forschungsgemeinschaft (DFG, German Research Foundation) under Germany's Excellence Strategy - EXC-2094 - 390783311. 
SKG and AS are supported by the Swiss National Science Foundation via the Grant No. PCEFP2\_181157. FM and REA acknowledge the support of the ERC-StG number
716151 (BACCO). The simulations were analysed using \textsc{Tools21cm} \citep[][]{giri2020tools21cm}.
%All the analysis and figures are made using the following packages: \textsc{numpy} \citep[][]{van2011numpy}, \textsc{scipy} \citep[][]{virtanen2020scipy}, \textsc{Tools21cm} \citep[][]{giri2020tools21cm} and \textsc{matplotlib} \citep[][]{hunter2007matplotlib}.

\end{acknowledgements}

% WARNING
%-------------------------------------------------------------------
% Please note that we have included the references to the file aa.dem in
% order to compile it, but we ask you to:
%
% - use BibTeX with the regular commands:
\bibliographystyle{aa} % style aa.bst
\bibliography{reference} % your references Yourfile.bib
%
% - join the .bib files when you upload your source files
%-------------------------------------------------------------------

\end{document}